# Long-wave models of thin film fluid dynamics

A.J. Roberts*

26 Sept 94

## Abstract

Centre manifold techniques are used to derive rationally a description of the dynamics of thin films of fluid. The derived model is based on the free-surface $\eta(x,t)$ and the vertically averaged horizontal velocity $\bar{u}(x,t)$. The approach appears to converge well and has significant differences from conventional depth-averaged models.

## 1 Introduction

Consider the dynamics of an incompressible viscous fluid flowing down an inclined plane of angle $\theta$. Long wave, low Reynolds number dynamics of the fluid film may be described by a well known equation [3, eq.(1)]. In the limit of small departures from a flat interface, it reduces to a Kuramoto-Sivashinsky equation [3, eq.(2)] or [4, eq.(20)]. Such long-wave models may be put on a firm theoretical footing with the aid of centre manifold techniques as has been done for Taylor dispersion[9, 19], flow reactors[1] and beam theory[17]. However, in many applications these particular long-wave models of fluid films have limited usefulness and a two-mode long-wave model of the dynamics, expressed in terms of the film thickness $\eta$ and mean velocity $\bar{u}$, is preferred [3, p110]. This note shows how to use centre manifold techniques to derive such two-mode long-wave models.

An advantage in using centre manifold techniques is that in a range of applications there is rigorous support for the resulting low-dimensional models [2]. However, rigorous theory does not yet cover this application. Nonetheless, based upon a large spectra gap, Section 2, I argue that a useful two-mode model can be constructed. By modify ing the Navier-Stokes equations, Section 3, I fit the problem into the centre manifol formalism and seek an asymptotic solution. The leading-order model of the dynamics Section 4, exhibits significant quantitative differences with other models based upo depth-averaging (or equivalently depth integration); these differences persist in a mor careful examination of the convergence of the model, Section 5. Proceeding to second order in the analysis, Section 6, I derive a dynamical model which includes terms pro portional to the Reynolds number; these terms are entirely absent in depth-average equations. Consequently, I deduce that depth-averaging is qualitatively and quanti tatively unsound except perhaps for low Reynolds number flows. Instead, I propos this centre manifold approach for developing models that agree quantitatively with th original governing differential equations.

With the $x$-axis along and the $y$-axis perpendicular to the plane of a bed at angle $\theta$ the Navier-Stokes and continuity equations for the dynamics of a film of incompressibl fluid are:

$$\frac{\partial u}{\partial t} = -u\frac{\partial u}{\partial x} - v\frac{\partial u}{\partial y} + \nu\nabla^2 u - \frac{\partial p}{\partial x} + g\sin\theta\,, \qquad (1$$

$$\frac{\partial v}{\partial t} = -u\frac{\partial v}{\partial x} - v\frac{\partial v}{\partial y} + \nu\nabla^2 v - \frac{\partial p}{\partial y} - g\cos\theta\,, \qquad (2$$

$$0 = \frac{\partial u}{\partial x} + \frac{\partial v}{\partial y}\,. \qquad (3$$

with boundary conditions

$$\text{no-slip on } y=0: \quad u = v = 0 \qquad (4$$

$$\text{kinematic on } y=\eta: \quad \frac{\partial \eta}{\partial t} = v - u\frac{\partial \eta}{\partial x} \qquad (5$$

*Institut Non Linéaire de Nice, 1361 route des Lucioles, 06560 Valbonne, France. Permanent address: Dept. Maths & Computing, University of Southern Queensland, Toowoomba 4350, Australia. E-mail: `aroberts@usq.edu.au`

$$\text{normal stress on } y=\eta: \quad p = -\frac{\sigma\eta_{xx}}{(1+\eta_x^2)^{3/2}} + \\ + \frac{2\nu}{1+\eta_x^2}\left[\frac{\partial v}{\partial y} + \eta_x^2\frac{\partial u}{\partial x} - \eta_x\left(\frac{\partial u}{\partial y}+\frac{\partial v}{\partial x}\right)\right] \quad (6)$$

$$\text{tangential stress on } y=\eta: \quad (1-\eta_x^2)\left(\frac{\partial u}{\partial y}+\frac{\partial v}{\partial x}\right) + 2\eta_x\left(\frac{\partial v}{\partial y}-\frac{\partial u}{\partial x}\right) = 0 \quad (7)$$

where I have scaled quantities so that fluid density $\rho = 1$, and typical magnitudes of fluid thickness and horizontal velocity are of size 1. Thus, in terms of a typical velocity $U$ and a typical film thickness $H$, the dimensionless constants appearing in the above equations are as follows:

- $g$ is the reciprocal of the Froude number $1/Fr = gH/U^2$;
- $\nu$ is the reciprocal of the Reynolds number $1/Re = \mu/HU\rho$;
- and $\sigma$ is the Weber number $We = T/\rho U^2 H$.

There is a one-parameter family of equilibria of the governing equations (1–7), they correspond to uniform shear flow. In terms of a constant $h$ an equilibria is $\eta = h$, $p = -g\cos\theta(h-y)$, $u = \frac{g\sin\theta}{\nu}(hy - \frac{1}{2}y^2)$, and $v = 0$. Using centre manifold techniques we can construct low-dimensional dynamical models based on these equilibria. Note that because there is a family of equilibria, essentially parameterised by the free-surface height $\eta$, the derived model will permit order 1 changes in $\eta$ over the flow domain—just provided that the surface variations are slow.

## 2 The spectrum

In the absence of any $x$ variation, the spectrum of perturbations to any one of these equilibria is $\{\lambda_i\}$ where:

- $\lambda_0 = 0$ corresponds to the freedom to vary the fluid depth $h$;
- $\lambda_n = -\nu\left(\frac{\pi(2n-1)}{2h}\right)^2$ for $n = 1, 2, \ldots$ corresponding to viscously dissipating modes of horizontal shear $\tilde u_n = \sin\left[\frac{\pi(2n-1)}{2h}y\right]$.

The 0 eigenvalue permits the construction of a centre manifold model of flows which vary slowly in space[13]. Since, in the absence of spatial variations, there is one 0 eigenvalue, then the model would be written in terms of one slowly-varying mode. The evolution equation of such a model would be the traditional long-wave equation [3, eq.(1)] for th fluid depth $\eta$ (possibly with higher-order modifications).

However, there is also a clear spectral gap between $\lambda_0 = 0$ allied with $\lambda_1 = -\nu\left(\frac{\pi}{2h}\right)$ and the other eigenvalues headed by $\lambda_2 = -9\nu\left(\frac{\pi}{2h}\right)^2$. Hence a two-mode model base on the dynamics of $\eta$ and the gravest horizontal shear mode $\tilde u_1$ should be of interest i many applications [15]. The model should resolve accurately dynamics and transient on time-scales slower that $1/\lambda_2$.

## 3 The centre manifold analysis

I make the gravest mode, $\tilde u_1 = \sin\left(\frac{\pi y}{2\eta}\right)$, of the horizontal momentum equation (1) critical mode by modifying (1) to

$$\frac{\partial u}{\partial t} = -u\frac{\partial u}{\partial x} - v\frac{\partial u}{\partial y} + \nu\nabla^2 u - \frac{\partial p}{\partial x} + g\sin\theta + (1-\gamma)\nu\frac{\pi^2}{4\eta^2}u\,. \quad (8$$

This modification to the right-hand side shifts the eigenvalues $\{\lambda_1, \lambda_2, \ldots\}$ by just th coefficient of the new term in $u$. Treating $\gamma$ as small in the asymptotic scheme (fo example, by the standard trick of appending $\partial\gamma/\partial t = 0$ to the set of equations, as i done to unfold bifurcations[2]), $\tilde u_1(y)$ corresponds to a 0 eigenvalue, and we construc a two-mode centre manifold model. Ultimately, I set $\gamma = 1$ to recover approximat expressions for the original problem. The model describes the long-term evolution o thin film dynamics in terms of the two critical modes $\eta$ and $\bar u\tilde u_1(y)$, slowly varying in and $t$.

Having identified the critical modes, the subsequent analysis is straightforward a explained in more detail for other systems elsewhere[13, 9, 17, for example]. Collectin the unknown fields into $\mathbf{u} = (u, v, p)$ we seek a low-dimensional centre manifold give by the asymptotic expansion

$$\mathbf{u}(x,y,t) = \mathbf{V}(\eta,\bar u,\gamma,y) \sim \sum_{m,n=0}^{\infty}\gamma^m\mathbf{V}^{m,n}(\eta,\bar u,y)\,,$$

where all the $x$ and $t$ dependence is solely through the evolution of $\eta(x,t)$ and $\bar u(x,t$ Thus we pose an evolution of

$$\frac{\partial\eta}{\partial t} = G_\eta(\eta,\bar u,\gamma) \sim \sum_{m,n=0}^{\infty}\gamma^m G_\eta^{m,n}(\eta,\bar u)\,,$$

$$\frac{\partial \bar{u}}{\partial t} = G_u(\eta, \bar{u}, \gamma) \sim \sum_{m,n=0}^{\infty} \gamma^m G_u^{m,n}(\eta, \bar{u}) .$$

In these asymptotic expansions the superscript $^{m,n}$ denotes a term which is of order $m$ in $\gamma$, as is explicitly shown, and implicitly of order $n$ in the bed-slope $\theta$ and spatial derivatives $\partial/\partial x$.[13] Further, it is convenient (although not essential) to consider the surface tension parameter $\sigma$ to be large, of order $-2$, so that surface tension effects are promoted to the leading order.

Substituting this ansatz into the governing differential equations (2–8), which may be written in the abstract form $\mathcal{K}\frac{\partial \mathbf{u}}{\partial t} = \mathcal{L}\mathbf{u} + \mathbf{N}(\mathbf{u})$ where $\mathcal{L}$ is the singular leading-order operator, we aim to solve

$$\mathcal{K}\left(\frac{\partial \mathbf{V}}{\partial \eta} G_\eta + \frac{\partial \mathbf{V}}{\partial \bar{u}} G_u\right) = \mathcal{L}\mathbf{V} + \mathbf{N}(\mathbf{V}) .$$

Then substituting the asymptotic expansions into this equation and equating terms of the same order gives a hierarchy, in order $m$ and $n$, of equations to solve. The equations are made well-posed by requiring that the "amplitudes" of the parameterisation of the model are precisely the free-surface height $\eta$ and the mean horizontal velocity

$$\bar{u}(x,t) = \frac{1}{\eta}\int_0^\eta u\, dy .$$

The resulting equations were solved and checked using the `reduce` computer algebra package.

Note that I apply only formal centre manifold *techniques*; the rigorous theory for this situation has not yet been developed. The difficulty primarily lies in the "infinite" dimension of the centre manifold—"infinite" because it is parameterised by continuous functions of $x$. For example, the theory of Gallay[6] on infinite dimensional centre manifolds requires that the nonlinear perturbations are bounded, but here the nonlinear terms, $\mathbf{N}(\mathbf{u})$, involve the spatial derivative $\partial/\partial x$ which is *unbounded*. I anticipate that in time theory will be developed to rigorous support these systematic techniques.

# 4 The leading order model

The leading order problem in the hierarchy of equations is (where superscripts $^{0,0}$ are omitted for clarity)

$$0 = -v\frac{\partial u}{\partial y} + \nu\frac{\partial^2 u}{\partial y^2} + \nu\frac{\pi^2}{4\eta^2}u , \quad 0 = -v\frac{\partial v}{\partial y} + \nu\frac{\partial^2 v}{\partial y^2} - \frac{\partial p}{\partial y} - g , \qquad 0 = \frac{\partial v}{\partial y} ,$$
$$u = v = 0 \text{ on } y = 0 , \qquad v = \frac{\partial u}{\partial y} = p + \sigma\eta_{xx} - 2\nu\frac{\partial v}{\partial y} = 0 \text{ on } y = \eta .$$

with the basic solution of a shear flow in fluid of depth $\eta$:

$$v^{0,0} = 0 , \quad p^{0,0} = g(\eta - y) - \sigma\eta_{xx} , \quad u^{0,0} = \bar{u}\frac{\pi}{2}\sin\left(\frac{\pi y}{2\eta}\right) . \qquad (9$$

Higher orders in the analysis show how space variations in $\eta$ and $\bar{u}$ couple to the physica processes in the Navier-Stokes equations to produce modifications of this leading orde structure together with time evolution of $\eta$ and $\bar{u}$.

For example, the next order adjustments to the pressure and velocity fields are, i terms of the scaled vertical coordinate $\zeta = \pi y/(2\eta)$,

$$\begin{aligned}
p^{0,1} &= -\frac{\pi\nu}{2}\bar{u}_x(\sin\zeta + 1) + \frac{\pi\nu}{2\eta}\bar{u}\eta_x\zeta\cos\zeta , \\
v^{0,1} &= -\eta\bar{u}_x(1-\cos\zeta) + \bar{u}\eta_x(\zeta\sin\zeta + \cos\zeta - 1) , \\
u^{0,1} &= \frac{1}{\nu}\bar{u}\eta^2\bar{u}_x\left(\frac{2}{\pi}\zeta\cos\zeta - \cos\zeta - \frac{\pi}{2}\sin\zeta + \frac{1}{2\pi}\zeta^2\sin\zeta + \frac{3}{\pi}\sin\zeta - \frac{1}{2}\zeta\sin\zeta + 1\right) \\
&\quad -\frac{1}{\nu}\bar{u}^2\eta\eta_x\left(+\frac{1}{6}\cos(2\zeta) - \frac{2}{3\pi}\zeta\cos\zeta + \frac{1}{3}\cos\zeta + \frac{\pi}{4}\sin\zeta - \frac{1}{2\pi}\zeta^2\sin\zeta - \right. \\
&\qquad \left. -\frac{5}{3\pi}\sin\zeta + \frac{1}{2}\zeta\sin\zeta - \frac{1}{2}\right) \quad + \left(\frac{\sigma}{\nu}\eta^2\eta_{xxx} - \frac{g\cos\theta}{\nu}\eta^2\eta_x + \frac{g\sin\theta}{\nu}\eta^2\right)\times \\
&\quad \times\left(\frac{4}{\pi^2}\cos\zeta - \frac{8}{\pi^3}\zeta\cos\zeta + \frac{2}{\pi}\sin\zeta - \frac{8}{\pi^3}\sin\zeta - \frac{4}{\pi^2}\right) .
\end{aligned}$$

The three structure functions appearing in $u^{0,1}$ are plotted in Figure 1 as a function of $\zeta$ Observe that they all have much the same shape that will tend to flatten the horizonta velocity profile $u^{0,0}(y)$ (say positive): in thicker flows due to the $\eta^2/\nu$ factor; if th free-surface slopes downwards due to the $-\eta_x$ factors, as is appropriate for a convergin flow, or if the mean velocity is increasing due to the $\bar{u}\bar{u}_x$ factor; or through surfac tension effects, $\propto \sigma\eta_{xxx}$, if the curvature increases in $x$. Indeed, this last effect ma predict the recirculation regions seen in Figure 7(b,c) from the numerical simulation of Chang *et al* [4], especially as the predicted effect increases with Reynolds numbe $(1/\nu)$. All of these dynamic modifications to the velocity profile are proportional to th Reynolds number.

As you can imagine, the formulae for the fields rapidly become more complicated a higher orders. I do not record any of these details here, instead I concentrate upon th evolution equations for the model's parameters, $\eta$ and $\bar{u}$.

As usual, the above adjustments are found by inverting the singular linear operato $\mathcal{L}$. This is only possible if the right-hand side is in the range of the singular operato and this solvability condition determines the evolution terms $G_u^{0,1}$ and $G_\eta^{0,1}$. Similarl at first order in $\gamma$ we determine $G_u^{1,0}$ and $G_\eta^{1,0}$. The only non-zero $G_\eta^{m,n}$ at any order i $G_\eta^{0,1}$, which gives the exact result

$$\frac{\partial \eta}{\partial t} = -\frac{\partial(\eta\bar{u})}{\partial x}\,, \qquad (10$$

at all orders. This is a direct consequence of conservation of fluid.

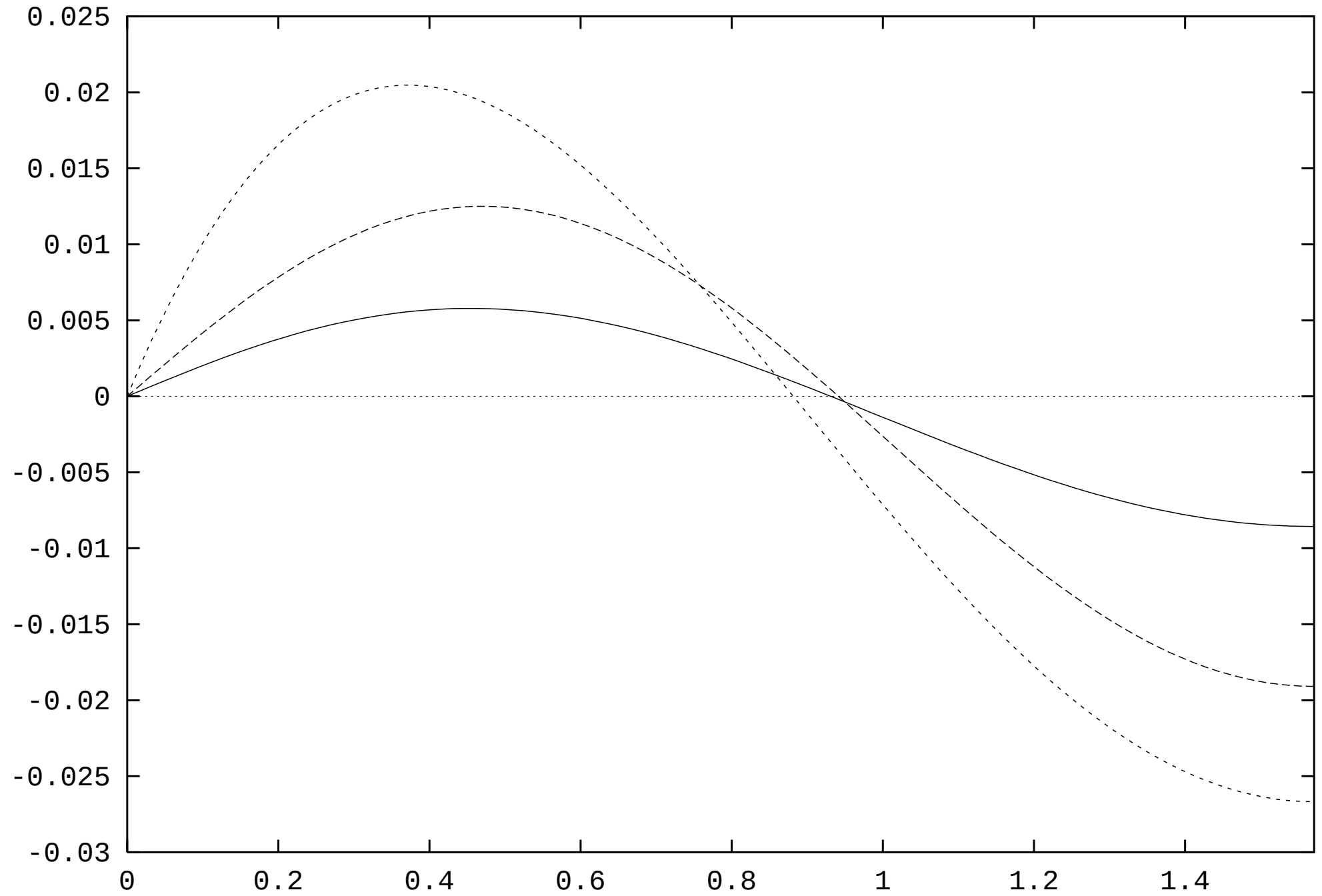

Figure 1: profiles of the three structure functions appearing in $u^{0,1}$, the first correction to the leading order horizontal velocity profile $u^{0,0}$, as a function of the scaled vertical coordinate $\zeta = \pi y/(2\eta)$: ——, the $\bar{u}\eta^2\bar{u}_x/\nu$ term; – – –, the $-\bar{u}^2\eta\eta_x/\nu$ term; - - - -, the remaining term.

Henceforth, I concentrate on the evolution equation for the mean horizontal velocit $\bar{u}$. The leading order evolution is governed by

$$\begin{aligned} \frac{\partial \bar{u}}{\partial t} &\sim G_u^{0,1} + \gamma G_u^{1,0} \\ &= \frac{8}{\pi^2}\left(g\sin\theta - g\cos\theta\frac{\partial\eta}{\partial x} + \sigma\frac{\partial^3\eta}{\partial x^3}\right) - \frac{3}{2}\bar{u}\frac{\partial\bar{u}}{\partial x} - \frac{\pi^2}{4}\frac{\nu\gamma}{\eta^2}\bar{u} - \frac{1}{6}\frac{\bar{u}^2}{\eta}\frac{\partial\eta}{\partial x} \qquad (11 \end{aligned}$$

upon setting $\gamma = 1$ to recover a description relevant to the original problem. Comparin the leading order coefficients with those from equation (19) of Prokopiou *et al* [11]:

- from the above $\bar{u}\bar{u}_x$ term the shape factor used in other approaches should be 5/4 whereas from the above $\bar{u}^2\eta_x/\eta$ term it should be 7/6—the value of 1.2 used b Prokopiou *et al* is perhaps a reasonable compromise;
- the coefficient of viscous decay, the $\nu\bar{u}/\eta^2$ term, is $\pi^2/4 \approx 2.46$, some 17% weake than the conventional 3;
- but a more significant difference is that gravity and horizontal pressure gradient are actually less effective than depth-averaging indicates (by nearly 20%, the facto $8/\pi^2 \approx 0.81$ instead of 1).

The first two differences could be attributed to the different "shape" of the leadin order velocity profile $\tilde{u}_1(y)$, here sinusoidal instead of parabolic. However, the last i fundamentally new. The phenomenon is due to the response of the fluid, primaril $\sin(\pi y/2/\eta)$, being at an angle to the forcing 1 (either due to gravity or horizonta pressure gradients) when considered in the space of functions on $[0,\eta]$. Consequentl the forcing is less effective.

A further comparison is made with the lubrication approximation, see equation (2) o Moriarty *et al* [10] for example. The lubrication approximation arises directly from (11 by assuming that the horizontal velocity is small and has viscous dissipation balance by production from the $x$ component of gravity and surface gradients. Thus

$$\bar{u} \approx \frac{32}{\pi^4}\frac{\eta^2}{\nu}\left(g\sin\theta - g\cos\theta\frac{\partial\eta}{\partial x} + \sigma\frac{\partial^3\eta}{\partial x^3}\right)\,.$$

The coefficient $32/\pi^4$ is within 1.5% of the 1/3 needed to agree with lubrication theory.

However, in order to obtain the leading order approximation (11), I introduced $\gamma$ to modify the original problem. It is necessary to re-examine the above comparisons in a more accurate solution of the original problem by proceeding to higher order in $\gamma$.

# 5 Convergence

Computations indicate that the dependence upon $\gamma$ converges very quickly; or at least is an asymptotic series with excellent low-order summability. Solving for terms up to order $\gamma^5$, I obtain the following for the evolution of $\bar{u}$:

$$
\begin{aligned}
\frac{\partial \bar{u}}{\partial t} &\sim \sum_{m=0}^{5} \gamma^m \left(G_u^{m,0} + G_u^{m,1}\right) \\
&= (.810569 + .013211\gamma - .001473\gamma^2 + .000178\gamma^3 - .000022\gamma^4 + .000002\gamma^5) \times \\
&\quad \times \left(g\sin\theta - g\cos\theta\frac{\partial \eta}{\partial x} + \sigma\frac{\partial^3\eta}{\partial x^3}\right) \qquad (12) \\
&\quad -(1/6 - .015112\gamma - .002723\gamma^2 - .000446\gamma^3 - .000069\gamma^4 - .000010\gamma^5)\frac{\bar{u}^2}{\eta}\frac{\partial \eta}{\partial x} \\
&\quad -(3/2 + .003634\gamma + .000440\gamma^2 + .000054\gamma^3 + .000006\gamma^4 + .000001\gamma^5)\bar{u}\frac{\partial \bar{u}}{\partial x} \\
&\quad -\frac{\pi^2}{4}\gamma\frac{\nu\bar{u}}{\eta^2}
\end{aligned}
$$

It seems reasonable to conclude from these expressions that the leading order in $\gamma$ is generally accurate to nearly two decimal places; whereas evaluating the above to terms in $\gamma^3$ for $\gamma = 1$ should give about four decimal place accuracy in the coefficients of the evolution equation.

I conjecture that the reason this expansion in $\gamma$ works so well here is that the gravest eigenmode is well represented by the critical mode in the modified equation; in fact they are identical. Thus, here the approximation in $\gamma$ is primarily an approximation in the dynamics. The approximation that $\lambda_1 \approx 0$ when compared with the neglected dynamics, headed by $\lambda_2$, is very good. Indeed, the rate at which the coefficients in the $\gamma$ expansions are decreasing is clearly about $\lambda_1/\lambda_2 = 1/9$.

However, earlier work on a simple dynamical system[12] has shown that we should expect high-order nonlinear terms to have poorer convergence. Generally expect that a term in $\bar{u}^n$ will involve the Taylor expansion of $1/(\lambda_2 - n\lambda_1\gamma)$ which will not converge at $\gamma = 1$ for terms of order $n \geq 9$. Thus although the low orders of the model converge well in $\gamma$, expect that it is only asymptotic in $\eta$ and $\bar{u}$ and that we should only use a truncation of less than ninth order. In practise this is not a significant restriction a only the first few orders are of interest.

An outstanding feature of the evolution equation (12) is that the quantitative dis crepancies with traditional depth-averaged models, such as equation (19) in [11], remai almost unaltered. However, the small discrepancy with lubrication theory disappears setting $\gamma = 1$ the shown coefficients in the first line of (12) sums to within roundoff er ror, $2 \times 10^{-6}$, of the $\pi^2/12$ needed for exact agreement. I deduce that depth-averagin is quantitatively unsound for these sorts of shear flows.

# 6 Second order model

Having established that there is very good convergence in the parameter $\gamma$, I exten the model to higher order in spatial gradients and slope. Using **reduce** I find that t third-order in $\gamma$ and to second-order in other quantities

$$
\begin{aligned}
\frac{\partial \bar{u}}{\partial t} \sim\ & 0.8225\left(g\sin\theta - g\cos\theta\eta_x + \sigma\eta_{xxx}\right) - 1.504\bar{u}\bar{u}_x - \frac{\pi^2}{4}\frac{\nu}{\eta^2}\bar{u} - 0.1484\frac{1}{\eta}\bar{u}^2\eta_x \\
& +\nu\left(-0.5834\frac{1}{\eta}\bar{u}\eta_{xx} - 0.1066\frac{1}{\eta^2}\bar{u}\eta_x^2 + 4.833\frac{1}{\eta}\eta_x\bar{u}_x + 4.093\bar{u}_{xx}\right) \\
& +\frac{g\cos\theta}{100\nu}\left(-1.231\eta^2\bar{u}\eta_{xx} - 2.368\eta\bar{u}\eta_x^2 - 0.04758\eta^2\eta_x\bar{u}_x + 0.4821\eta^3\bar{u}_{xx}\right) \\
& +\frac{g\sin\theta}{100\nu}\left(2.526\eta\bar{u}\eta_x + 0.7983\eta^2\bar{u}_x\right) \qquad (13 \\
& +\frac{\sigma}{100\nu}\left(1.231\eta^2\bar{u}\eta_{xxxx} + 1.896\eta\bar{u}\eta_x\eta_{xxx} - 0.7031\eta^2\bar{u}_x\eta_{xxx} - 0.4723\eta\bar{u}\eta_{xx}^2\right. \\
& \quad +4.004\bar{u}\eta_x^2\eta_{xx} - 1.889\eta\eta_x\bar{u}_x\eta_{xx} - 2.252\eta^2\eta_{xx}\bar{u}_{xx} - 3.859\frac{1}{\eta}\bar{u}\eta_x^4 \\
& \quad \left.+2.669\eta_x^3\bar{u}_x - 0.9445\eta\eta_x^2\bar{u}_{xx} - 1.501\eta^2\eta_x\bar{u}_{xxx} - 0.4821\eta^3\bar{u}_{xxxx}\right) \\
& +\frac{1}{100\nu}\left(-1.014\eta\bar{u}^3\eta_{xx} - 1.854\bar{u}^3\eta_x^2 - 0.02076\eta\bar{u}^2\eta_x\bar{u}_x + 0.7778\eta^2\bar{u}^2\bar{u}_{xx}\right. \\
& \quad \left.+0.1226\eta^2\bar{u}\bar{u}_x^2\right) .
\end{aligned}
$$

This is certainly a complicated evolution equation; which is one reason for not proceed ing to any higher order.[1] The striking qualitative differences between this equation an other proposed models are twofold.

[1] Another reason is that the algebraic details are sufficiently complicated so that the **reduce** prograr was restricted to this order by computational resources.

- Most obvious are all the terms proportional to the Reynolds number $(1/\nu)$ in the above equation (13); they have no counterpart in the depth-averaged models (such as [11, eq.(19)]). Such terms arise through the subtle interplay of nonlinear interactions and velocity shear with vertical diffusion (they are somewhat akin to the effective diffusion term of Taylor's description[18] of shear-enhanced dispersion in a channel)—they are very hard to derive without the systematic approach of centre manifold theory. Being proportional to the Reynolds number they are likely to be of interest in the faster flows, though the factor of 1/100 indicates that the Reynolds number will need to be comparable to 100 before these terms are significant.

- The other qualitative difference is the absence terms in $\sigma\eta\eta_x^2\eta_{xxx}$ and $\sigma\eta\eta_x\eta_{xx}^2$ as appear in [11, eq.(19)]. Such terms would arise at the next order in this analysis as they involve five spatial derivatives and $\sigma$ is of order $-2$. The centre manifold approach suggests that other terms are more important.

For a quantitative comparison of the other terms, with equation (19) in [11], I recast (13) in terms of the fluid flux $q(x,t) = \eta\bar{u}$.

$$
\begin{aligned}
\frac{\partial q}{\partial t} \sim\ & +0.8225\left(g\sin\theta\eta - g\cos\theta\eta\eta_x + \sigma\eta\eta_{xxx}\right) \\
& \quad +1.356\frac{1}{\eta^2}q^2\eta_x - 2.504\frac{1}{\eta}qq_x - \frac{\pi^2}{4}\frac{\nu}{\eta^2}q \\
& +\nu\left(-4.676\frac{1}{\eta}q\eta_{xx} + 3.459\frac{1}{\eta^2}q\eta_x^2 - 3.353\frac{1}{\eta}\eta_x q_x + 4.093q_{xx}\right) \\
& +\frac{g\cos\theta}{100\nu}\left(-1.713\eta^2q\eta_{xx} - 1.357\eta q\eta_x^2 - 1.012\eta^2\eta_x q_x + 0.4821\eta^3 q_{xx}\right) \\
& +\frac{g\sin\theta}{100\nu}\left(1.727\eta q\eta_x + 0.7983\eta^2 q_x\right) \\
& +\frac{\sigma}{100\nu}\left(1.713\eta^2q\eta_{xxxx} + 0.244\eta q\eta_x\eta_{xxx} + 1.225\eta^2q_x\eta_{xxx} - 1.113\eta q\eta_{xx}^2\right. \\
& \quad +10.68q\eta_x^2\eta_{xx} - 4.451\eta\eta_x q_x\eta_{xx} + 0.6404\eta^2\eta_{xx}q_{xx} - 10.98\frac{1}{\eta}q\eta_x^4 \\
& \quad \left. +7.12\eta_x^3 q_x - 2.225\eta\eta_x^2 q_{xx} + 0.4269\eta^2\eta_x q_{xxx} - 0.4821\eta^3 q_{xxxx}\right) \\
& +\frac{1}{100\nu}\left(-1.792\frac{1}{\eta}q^3\eta_{xx} - 0.1961\frac{1}{\eta^2}q^3\eta_x^2 - 1.78\frac{1}{\eta}q^2\eta_x q_x\right. \\
& \quad \left. +0.7778q^2q_{xx} + 0.1226qq_x^2\right) .
\end{aligned} \tag{14}
$$

The earlier discussed discrepancies remain. In addition, instead of the terms

$$\nu\left(5q_{xx} + \frac{6q\eta_x^2}{\eta^2} - \frac{6q_x\eta_x}{\eta} - \frac{6q\eta_{xx}}{\eta}\right),$$

the centre manifold analysis recommends (to one decimal place)

$$\nu\left(4.1q_{xx} + \frac{3.5q\eta_x^2}{\eta^2} - \frac{3.4q_x\eta_x}{\eta} - \frac{4.7q\eta_{xx}}{\eta}\right).$$

Using these new coefficients should improve agreement between the one-dimensional model and the full Navier-Stokes equations.

# 7 Remarks

Equation (13) (or equivalently (14)), in conjunction with the continuity equation (10) form a consistent one-dimensional model for the dynamics of a fluid film. Such centre manifold models are naturally of mixed order and so there is a great deal of flexibility in choosing which terms to retain in any given application of the model—a decision that can be delayed until application, and based upon the particular parameters of the application, rather than forced by the restrictive scaling principles of primitive mathematical methods such as that of multiples scale. Also, the geometric picture of centre manifold theory will lead to, in further work, expressions for the correct modelling of initial conditions [14], forcing [5] and boundary conditions [16]. Here I just focused on a derivation of the evolution equations.

Due to isotropy, the generalisation of equations (13) and (14) to the dynamics of a thin sheet of fluid is straightforward: the mean horizontal velocity and horizontal derivatives just turn into the corresponding vector velocity and gradient; and the gravity factors become appropriate dot products.

A very similar approach to the one taken here could be used in a wide range of physical problems such as the dynamics of slender jets [7], roll waves in muddy fluid [8] and long waves on turbulent flow (in preparation).

**Acknowledgements** I thank H.C. Chang and L. Schwartz for their encouragement in this work, and the Institut Non Linéaire de Nice for hospitality during its preparation. The research is assisted by grants from the Australian Research Council.